\RequirePackage{fix-cm}
\documentclass[smallextended]{svjour3}       
\smartqed  
\usepackage[a4paper]{geometry}
\usepackage[affil-it]{authblk}
\usepackage{amssymb}
\usepackage[T1]{fontenc}
\usepackage[english]{babel}
\usepackage{hyperref}
\usepackage{natbib}
\usepackage{epsfig}
\usepackage{chngpage}
\usepackage{amsmath}
\usepackage{latexsym}
\usepackage{longtable}
\usepackage{graphicx}
\usepackage{graphics}
\usepackage{booktabs}
\usepackage[singlelinecheck={true},small,bf]{caption}
\usepackage{subfig}
\usepackage{float}
\usepackage{wrapfig}
\usepackage{pdflscape}
\usepackage{pdfpages}
\usepackage{txfonts}
\usepackage{textcomp}

\def \cgssb {{\rm\,erg\,s^{-1}\,cm^{-2}\,arcsec^{-2}}}

\begin{document}

\title{Unveiling the faint ultraviolet Universe
}


\author{
A. Zanella$^{1,2}$         \and
C. Zanoni$^2$ \and
F. Arrigoni-Battaia$^3$ \and
A. Rubin$^2$ \and
A. F. Pala$^2$ \and
C. Peroux$^{2,4}$ \and
R. Augustin$^5$ \and
C. Circosta$^{2,6}$ \and
E. Emsellem$^{2,7}$ \and
E. George$^2$ \and
D. Milakovi\'c$^2$ \and
R. van der Burg$^2$ \and
T. Kupfer$^8$
}

\authorrunning{A. Zanella, C. Zanoni, F. Arrigoni-Battaia et al.} 

\institute{$^1$Istituto Nazionale di Astrofisica, Vicolo dell'Osservatorio 5, 35122 Padova (Italy) - Tel.: +39 0498 293507 -  \email{anita.zanella@inaf.it}     
           \and
           $^2$European Southern Observatory, Karl-Schwarzschild-Str. 2,
           85748 Garching bei M\"unchen (Germany)
           \and
           $^3$ Max-Planck-Institut f\"ur Astrophysik, Karl-Schwarzschild-Str 1, 85748, Garching bei M\"unchen, Germany
           \and
           $^4$ Aix Marseille Universit\'e, CNRS, LAM (Laboratoire d'Astrophysique de Marseille) UMR 7326, 13388, Marseille, France
           \and
           $^5$ Space Telescope Science Institute, 3700 San Martin Drive, Baltimore, MD, 21218, USA
           \and
           $^6$ Department of Physics \& Astronomy, University College London, Gower Street, London WC1E 6BT, United Kingdom
           \and
           $^7$ Univ. Lyon, Univ. Lyon1, ENS de Lyon, CNRS, Centre de Recherche Astrophysique de Lyon UMR5574, F-69230 Saint-Genis-Laval (France)
           \and
           $^8$ Kavli Institute for Theoretical Physics, University of California, Santa Barbara, CA 93106, USA
}

\date{Received: date / Accepted: date}

\maketitle

\begin{abstract}
With this paper we participate to the call for ideas issued by the European Space Agency to define the Science Program and plan for space missions from 2035 to 2050. In particular we present five science cases where major advancements can be achieved thanks to space-based spectroscopic observations at ultraviolet (UV) wavelengths. We discuss the possibility to (1) unveil the large-scale structures and cosmic web in emission at redshift $\lesssim 1.7$; (2) study the exchange of baryons between galaxies and their surroundings to understand the contribution of the circumgalactic gas to the evolution and angular-momentum build-up of galaxies; (3) constrain the efficiency of ram-pressure stripping in removing gas from galaxies and its role in quenching star formation; (4) characterize the progenitor population of core-collapse supernovae to reveal the explosion mechanisms of stars; (5) target accreting white dwarfs in globular clusters to determine their evolution and fate.
These science themes can be addressed thanks to UV (wavelength range $\lambda \sim 90 - 350$ nm) observations carried out with a panoramic integral field spectrograph (field of view $\sim 1 \times 1$ arcmin$^2$), and medium spectral (R = 4000) and spatial ($\sim 1" - 3"$) resolution. Such a UV-optimized instrument will be unique in the coming years, when most of the new large facilities such as the Extremely Large Telescope and the James Webb Space Telescope are optimized for infrared wavelengths.
\end{abstract}

\section{Introduction}
\label{intro}
The European Space Agency (ESA) has started planning the next cycle of space missions by establishing the long-term ESA Science Program Voyage 2050. This follows the current plan (Cosmic Vision, extending up to 2035) and is the framework in which ESA space missions from 2035 up to 2050 will be defined. In order to keep the Science Program a bottom-up process and gather inputs from the scientific community about the science themes that should be covered by the Voyage 2050 planning cycle, ESA has issued a call for White Papers in March 2019. In this paper we discuss five science cases, spanning different fields of astrophysics, presenting scientific challenges that can be addressed with a space-based mission in 2035 -- 2050, conducting spectroscopic observations in the ultraviolet (UV), a wavelength regime that is not accessible from the ground. 

\begin{itemize}
\item By detecting the intergalactic medium in emission it will be possible to directly unveil the cosmic web, whose existence is predicted by current theories of structure formation. This will enable studies of the exchange of baryons between galaxies (and quasars) and their surroundings, unveiling how the halo gas contributes to the evolution of galaxies and what mechanisms drive the galaxy angular momentum build-up through cosmic time. Finally, multiple detections of the intergalactic medium in emission will provide measurements of the UV background, a critical quantity used in simulations of galaxy formation, yet still poorly constrained in observations (Section \ref{subsec:cgm_fab}, \ref{subsec:cgm_celine}).

\item Observations of the neutral gas distribution (by mapping the Lyman-$\alpha$ emission) in low-redshift galaxy cluster members will clarify the efficiency with which ram-pressure stripping removes the gas from galaxies and the role of the environment in quenching star formation. These observations will be crucial to understand how and when the red sequence of galaxies is assembled in dense environments. These observations will be key in interpreting high redshift observations, where currently the Lyman-$\alpha$ is more easily accessible (Section \ref{subsec:ram_pressure}).

\item By observing statistical samples of supernovae in the UV it will be possible to characterize the progenitor population of core-collapse supernovae, providing the initial conditions for explosion models and allowing the community to progress in the understanding of the explosion mechanism of stars, as well as the final stages of stellar evolution (Section \ref{subsec:transients}). 

\item By targeting populations of accreting white dwarfs in globular clusters it will be possible to constrain the evolution and fate of these stars and investigate the properties of the most compact systems with the shortest orbital periods which are expected to be the brightest low-frequency gravitational wave sources. The possibility will also be explored that accreting white dwarfs are progenitors of Type Ia supernovae, which are fundamental sources to constrain cosmological distances and the current models for dark energy (Section \ref{subsec:binaries}).
\end{itemize}

A UV-optimized telescope (wavelength range $\lambda \sim $ 90 - 350 nm), equipped with a panoramic integral field spectrograph with a large field of view (FoV $\sim$ 1 $\times$ 1 arcmin$^2$), with medium spectral (R $= 4000$) and spatial ($\sim$ 1" -- 3") resolution will allow the community to simultaneously obtain spectral and photometric information of the targets, and tackle the science questions presented in this paper (Section \ref{sec:performances}).
The information-rich nature of the datasets provided by such an integral field spectrograph will represent a resource with considerable legacy value for the scientific community. Additionally, these observations will open up completely new areas of the parameter space, allowing the proposed mission to have a great potential for serendipitous discoveries. 

In the coming years, when most of the new large facilities such as the
Extremely Large Telescope (ELT) and the James Webb Space Telescope
(\textit{JWST}) will focus on the infrared (IR) wavelength range, and
the Hubble Space Telescope (\textit{HST}) will not be operational
anymore, a mission in the UV with the capability of observing
spectroscopically large areas of the sky will be unique. In synergy
with the Atacama Large Millimeter Array (ALMA), the ELT instruments, and Square Kilometer Array (SKA), but also with other space-based missions such as the Advanced Telescope for High Energy Astrophysics (\textit{Athena}) and the Laser Interferometer Space Antenna (\textit{LISA}) it will allow us to push further our current understanding of the Universe.

We present the main science themes to be addressed in the coming decades in
Section \ref{sec:sc}; we compare the characteristics of the proposed straw-man mission and instrument with other UV space missions in Section
\ref{sec:uniqueness} and its synergies with future facilities in Section~\ref{sec:synergies}; finally in Section
\ref{sec:performances} we sketch the high-level technical characteristics of the proposed instrument and address potential technological challenges.

\section{Science cases}
\label{sec:sc}
\subsection{Unveiling large-scale structures in emission at $z\lesssim1.7$}
\label{subsec:cgm_fab}

The current paradigm of large-scale structure formation predicts the presence of an intricate net of gaseous filaments connecting galaxies (e.g., \citealt{White1987}, \citealt{Bond1996}). The existence of this cosmic web, also 
known as intergalactic medium (IGM; \citealt{meiksin09}), is until now confirmed only indirectly by observations of the large-scale structures traced with galaxy surveys at low redshift and by studies of the 
Lyman-$\alpha$ (Ly$\alpha$) forest in absorption against background quasars.
Direct imaging of the cosmic web, probing its properties and evolution through cosmic history, will represent a major breakthrough for cosmology.
Directly detecting the IGM in this way, is predicted to be very challenging (surface brightness in Ly$\alpha$ predicted to be SB$_{\rm Ly\alpha}\sim10^{-19}-10^{-20}\cgssb$; \citealt{GW96,Bertone2012,Witstok2019}) because of the expected low densities for such gas ($n_{\rm H}\lesssim0.01$~cm$^{-2}$) and the budget of ionizing photons in the ultraviolet background (UVB; e.g., \citealt{hm12}).
A direct detection of the IGM appears to be so far elusive even with top-notch current facilities on 10m class telescopes (e.g., \citealt{Gallego2018,Wisotzki2018}), such as the Multi Unit Spectroscopic Explorer (MUSE; \citealt{Bacon2010}) and the Keck Cosmic Web Imager (KCWI; \citealt{Morrissey2012}). Ground-based instruments can target the Ly$\alpha$ emission only above the atmospheric cut-off ($z\gtrsim1.7$) and thus fight against a strong cosmological surface brightness dimming that scales as $(1+z)^{-4}$.

Astronomers have tried to bypass these limitations by searching the IGM signal around quasars. 
A quasar is expected to act as a flashlight, photoionizing the surrounding medium out to large distances. 
The ionized gas would then recombine, emitting as the main product hydrogen Ly$\alpha$ photons in copious amounts (e.g. \citealt{Rees1988, hr01}). 
The Ly$\alpha$ glow around quasars should then be boosted up to SB$_{\rm Ly\alpha}>10^{-19}\cgssb$, and therefore be within the reach of state-of-the-art instruments (\citealt{Cantalupo2005,kollmeier10}).

\begin{figure}
\centering
\includegraphics[width=\textwidth]{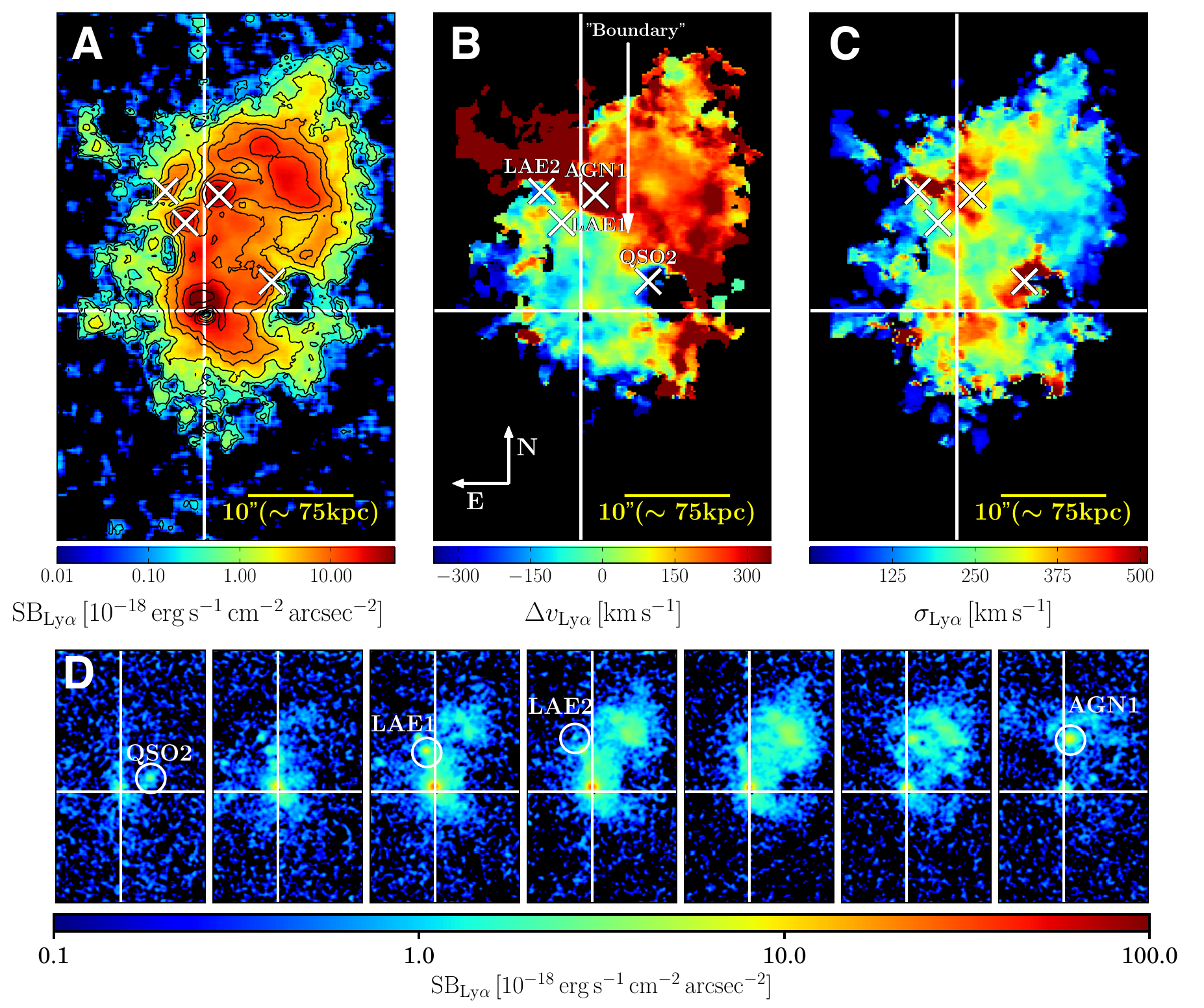}
\caption{An example of current large-scale structures detected in
  Ly$\alpha$ around quasars with VLT/MUSE at high redshift: the
  enormous Ly$\alpha$ nebula (ELAN) around the $z=3.164$ quasar SDSSJ~1020+1040 (figure adapted from \citealt{FAB2018}). 
{\bf (A)} ``optimally-extracted'' Ly$\alpha$ surface brightness map obtained
after subtraction of the quasar point-spread-function and continuum. 
The black contours indicate the isophotes corresponding to a signal-to-noise ratio of ${\rm S/N}=2,\,4,\,10,\,20,\,30,\,50$, and $100$. 
This image reveals an extremely bright nebula (SB$_{\rm Ly\alpha}\sim 10^{-17}\cgssb$) extending on the NW side of the quasar. 
Additional four strong Ly$\alpha$ emitters (diagonal crosses) are associated with the quasar, and the nebular emission. Two of these sources have been 
spectroscopically confirmed as AGN, making this system the third known quasar triplet at high $z$.
{\bf (B)} flux-weighted velocity-shift map with respect to 
the systemic redshift of the quasar obtained from the first-moment of the flux 
distribution.  A velocity shear between the SE and NW portion of the nebula is evident. The transition region is referred to as the ``Boundary''.
{\bf (C)} velocity dispersion map obtained from the second-moment of the 
flux distribution. Regions of higher dispersion 
($\sigma_{\rm Ly\alpha}\approx430$~km~s$^{-1}$) 
are visible in proximity of the three AGN, but overall the Ly$\alpha$ nebula shows 
quiescent kinematics ($\sigma_{v} < 270$~km~s$^{-1}$).
{\bf (D)} Each cut-out image (same size as A, B, and C) shows the surface brightness map of the ELAN within a  3.75~\AA\ layer ($3\times$ MUSE sampling) in the wavelength range  5058~\AA~$\lesssim \lambda \lesssim$~5084~\AA\ (from left to right). In all of the panels (A, B, C, D) the large white cross indicates the position of the 
quasar prior to PSF subtraction. 
Currently, similar large-scale emission cannot be probed at low $z$ ($z\lesssim 1.7$) because of the absence of appropriate facilities.} 
\label{fig:cgm_Fabrizio}
\end{figure}

Following this idea, several works targeted extended Ly$\alpha$ emission around high-redshift quasars
to constrain the physical properties of the diffuse gas phases out to intergalactic scales around individual objects (\citealt{TPW_2017}, \citealt{HuCowie1987,heckman91a,Moller2000b,Weidinger04,Weidinger05,Christensen2006,cantalupo14,martin14a,fab+16,Farina2017,Farina2019}).
For example, at $z\sim3$, it is now possible with MUSE to easily ($\sim 1$~hour on source; surface brightness limit of SB$_{\rm Ly\alpha}^{1~{\rm arcsec^2}}\sim10^{-18}\cgssb$) uncover the emission within 50 projected kpc from the targeted quasar, and to detect it up to distances of $\sim 80$~projected kpc (\citealt{Borisova2016,FAB2019}).
Notwithstanding these achievements when targeting individual quasars, it is evident that detections of diffuse emission at intergalactic distances ($>100$~kpc) at high $z$ are favored when additional active companions are present in close proximity (\citealt{hennawi+15,FAB2019,FAB2018}), or much more sensitive observations are conducted ($>10$~hours). 
Recent studies in the literature started to show new approaches in unveiling the IGM emission, passing from the
observations of individual quasars to (i) short (\citealt{Cai2018, FAB2019b}) or extremely long integrations ($\gg 40$~hours; \citealt{Lusso2019}) of multiple high-redshift quasars, or overdensities hosting quasars (\citealt{Cai2016}), and (ii) stacking of ultra deep observations of several galaxies (\citealt{Gallego2018,Wisotzki2018,Leclercq2020}). 
At face value, the detection of large-scale gas in emission still relies on the presence of active galaxies. 

In spite of the aforementioned difficulties in observing the IGM, state of the art integral field unit (IFU) spectrographs, by pushing the sensitivity of observations to SB$_{\rm Ly\alpha}\sim10^{-19}-10^{-20}\cgssb$, started to open new opportunities for the study of the gas distributed over large scales at high redshift. 
To study the large scale structures at the more accessible low-$z$ Universe, a new-generation space mission optimized for UV observations is needed to complement the ongoing efforts at high redshift.
By routinely detecting the IGM in emission at $z \lesssim 1.7$, this instrument will allow us to achieve the following science goals:

\begin{itemize}
\item{Connect the dots at low redshift: test our current view of the matter distribution at low $z$ by detecting the cosmic web in emission (through rest-frame UV emission lines) surrounding galaxies and quasars. This is crucial in our understanding of structure formation as we can currently only rely on galaxies as tracers of the distribution of large-scale structures at low $z$ \citep{Malavasi2017}.}
\item{Directly study the properties (e.g., density, metallicity) of the IGM in emission, complementing the information acquired from studies of the Ly$\alpha$ forest.}
\item{Provide measurements of the UV background at low $z$ thanks to multiple detection of the IGM in emission, building up independent constraints from the statistics of the IGM in absorption. The UV background is a critical quantity used in all simulations and models of structure formation, but it is still poorly constrained observationally. Its precise determination is key for our understanding of galaxy formation and evolution (e.g., \citealt{KhaireSrianand20192019}, \citealt{Faucher-Giguere2020}).}
\item{Study of the gas kinematics within the large-scale structures surrounding galaxies, allowing a direct characterization of the galaxy angular momentum build-up through cosmic time.}
\end{itemize}

\subsubsection{The need for space-based UV observations}
Recent works show that the most efficient and
effective way to detect emission extending to hundreds of kpc scales
around high-$z$ quasars and galaxies is the use of wide-field
IFU instruments (e.g., Figure \ref{fig:cgm_Fabrizio}; \citealt{FAB2018}). Space-based, wide-field spectroscopic observations in the wavelength range $\lambda \sim 90 - 350$ nm will allow us to achieve our scientific goals. Specifically, a space-based instrument is needed to observe the Ly$\alpha$ transition at low $z$ ($z\lesssim1.7$), where the cosmological surface brightness dimming is less severe. Assuming similar properties for the gas, it will be possible to target sources with $16\times$ lower surface brightness at $z \sim 0.5$ with respect to $z\sim3$. 
Furthermore, a wide field of view (1 $\times$ 1 arcmin$^2$, which corresponds to hundreds of kpc at $z \lesssim 1.7$) is needed to efficiently image the large-scale structures that subtend large areas on the sky. As we are interested in detecting large scale structures with very low surface brightness, large pixels (e.g. $\sim 1$~arcsec) are preferred to enhance sensitivity (as in e.g. KCWI; \citealt{Morrissey2012}). 

\subsection{Probing the emission of the Circumgalactic Medium around galaxies}
\label{subsec:cgm_celine}

Understanding the complex mechanisms regulating galaxy formation is
one of the main questions today in cosmology and astrophysics. The
question of how galaxies gather gas to sustain star formation is of
particular interest, as it sheds light on the fact that the star
formation rate (SFR) has been declining from $z \sim 2$ while diffusely
distributed hydrogen is still the dominant component for the total
baryonic mass budget (as compared to hydrogen in stars, \citealt{Madau2014}). The outflowing and accreting gas interacts around galaxies on scales up to hundreds of kpc (the Circumgalactic Medium, CGM). Studying the CGM is fundamental to understand the cosmic baryon cycle (\citealt{Steidel2010},
\citealt{Shull2014}, \citealt{TPW_2017}, \citealt{Peroux2020}) and it will provide
key constraints on the question of galaxy formation and
evolution. Absorption spectroscopy has already provided insights on the distribution and the chemical composition of the CGM gas from
a statistical point of view, given that typically only one line of sight per
galaxy can be studied due to the scarcity of background quasars in the
vicinity of galaxies (\citealt{Noterdaeme2012}, \citealt{Pieri2014},
\citealt{Quiret2016}, \citealt{Rahmani2016}, \citealt{Krogager2017},
\citealt{Augustin2018}, \citealt{Hamanowicz2020}). Therefore, mapping the CGM in emission is the
important next step to reach a full understanding of these complex regions.

\begin{figure}
\centering
\includegraphics[width=0.7\textwidth]{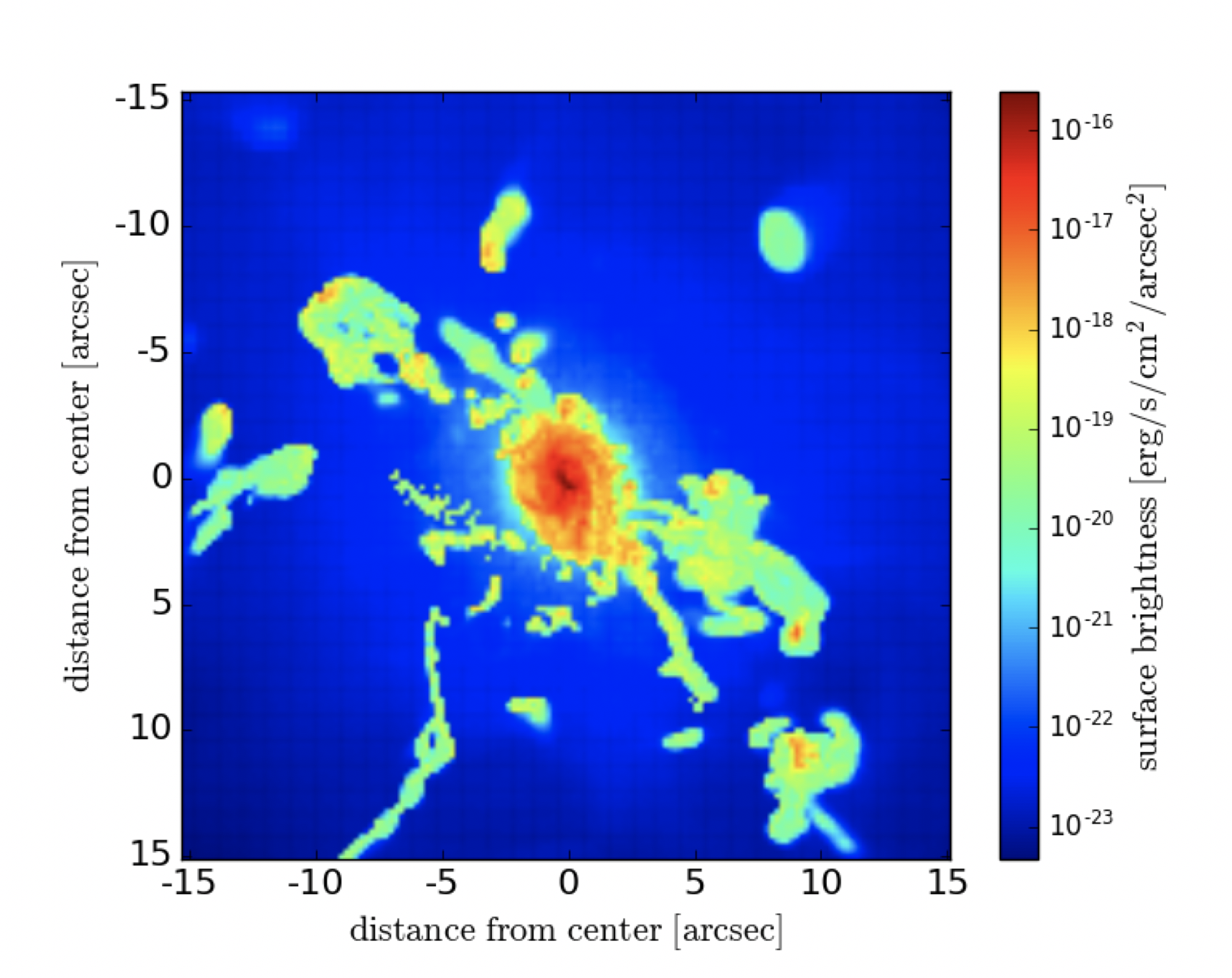}
\caption{Example of a mock Ly$\alpha$ CGM halo around a $z = 0.7$ galaxy, from RAMSES hydrodynamical cosmological simulations \citep{Augustin2019}. The Ly$\alpha$ line is at 206 nm~ and 1" $\sim$ 7 kpc at the redshift of this source. The CGM surface brightness levels shown in this map will be within reach of the proposed UV instrument.}
\label{fig:cgm_celine}
\end{figure}

\subsubsection{The need for space-based UV observations}
To achieve the science goals presented above it is key to detect and map different lines
(e.g. Ly$\alpha$, CIV, OVI, CVI, OVIII) arising from the CGM of low-redshift
galaxies. The signal (surface brightness) scales with $(1+z)^{-4}$ so that
lower redshift observations are considerably easier in turn requiring
space-born UV facilities to measure rest-frame UV lines. IFU-like capabilities are key for obtaining maps and kinematic reconstructions of the gas in the halos of galaxies. A field of view of
$\sim$ 1 $\times$ 1 arcmin$^2$ will cover most of the CGM region of a galaxy at
$z\sim1$. Modest spatial resolution (to increase sensitivity) and spectral
resolution (R $\sim$ 4000) are sufficient. Such a program will be complementary
to similar high-redshift projects that make use of extremely large
ground-based telescopes (i.e. ELT/HARMONI). An example of mock Ly$\alpha$ CGM
halo at $z=0.7$ from dedicated RAMSES cosmological hydrodynamical
simulations is shown in Figure \ref{fig:cgm_celine} \citep{Augustin2019}.

\subsection{Ram-pressure stripping and quenching in galaxy clusters}
\label{subsec:ram_pressure}

The existence of a well defined separation between massive, red,
early-type, quiescent galaxies and blue, late-type, actively
star-forming objects is a key leverage for the current modeling of galaxy formation and evolution. Nowadays, ``normal''
galaxies are thought to assemble their mass through a secular process
of star formation in a relatively steady state, forming a ``Main
Sequence'' up to high redshift (e.g. \citealt{Daddi2007},
\citealt{Speagle2014}) and following a tight gas-star formation rate
(SFR) density relation (``KS'' relation, \citealt{Schmidt1959},
\citealt{Kennicutt1998}). Deviations from this dynamic equilibrium may
occur in short starbursting events (typically associated with major mergers, \citealt{Sanders1996}) or due to the
cessation of star formation (``quenching''). Both these deviations
from equilibrium are poorly understood: why do galaxies suddenly
ignite the formation of thousands of stars per year? Why do they stop
forming stars? The deviations from the Main Sequence and the KS relation might be connected: the merger of gas-rich objects may first result in a burst of star formation, followed by a drop of the SFR and the subsequent quenching of the galaxy. The results of this process might be the compact, quiescent galaxies observed at $z \lesssim 2$ (e.g. \citealt{Cimatti2008}, \citealt{Toft2014}). However, it is still debated what is the mechanism responsible for stopping star formation and how galaxies are maintained quiescent for several Gyrs.

An additional piece of the puzzle is the correlation between galaxy color (and
morphology) with both local environment and stellar
mass (\citealt{Dressler1980}, \citealt{Bell2004}, \citealt{Peng2010}).
The environmental dependence seems to indicate that not only internal, but also external, physical processes play a role in shaping the star formation of
galaxies at all cosmic ages (\citealt{Boselli2006},
\citealt{Blanton2009}). Internal mechanisms such as feedback from supernovae and active galactic nuclei (AGN, e.g. \cite{Hopkins2012},
\citealt{Croton2006}), dynamical stabilization (\citealt{Martig2009}), and gravitational heating \citep{Johansson2009} are deemed responsible for suppressing and quenching
star formation at all densities. Environmental processes in the form
of ram pressure or viscous stripping, tidal interactions
\citep{Gunn1972}, and the consumption of the galaxy's gas reservoir by star formation without further replenishment from the cosmic web because of accumulation of hot plasma inducing shocks and heating up infalling gas (e.g. \citealt{Larson1980}, \citealt{Peng2015}; 
\citealt{Dekel2006}) seem to be key players in shaping the observed color bimodality
within galaxy clusters and groups, especially at the faint end of the galaxy
luminosity function.

\begin{figure}
\centering
\includegraphics[width=0.9\textwidth]{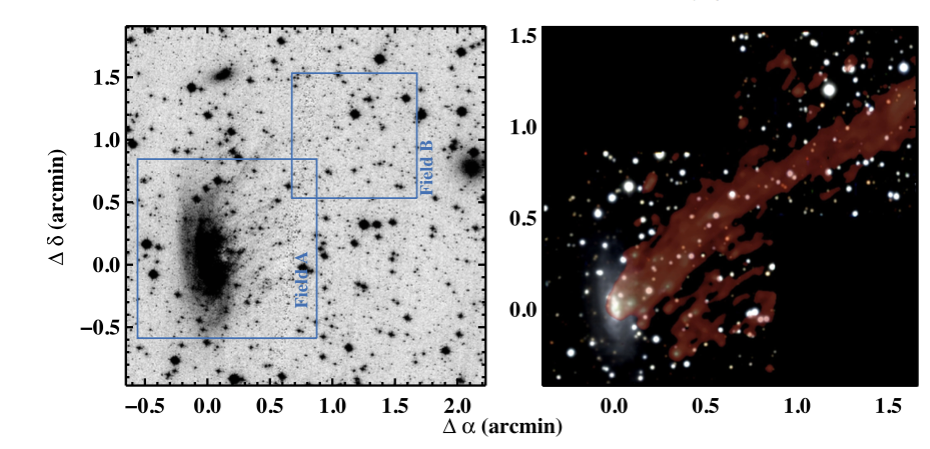}
\caption{Example of a local cluster member undergoing ram pressure stripping observed with VLT/MUSE: ESO137-001 at $z = 0.016$ \citep{Fumagalli2014}. \textbf{Left}: HST/ACS image in the F475W filter with, superposed, the MUSE field of view at the two locations targeted by the observations. \textbf{Right}: RGB color image obtained combining images extracted from the MUSE data cube in three wavelength intervals ($\lambda = 500 - 600$ nm~ for the B channel, $\lambda = 600 - 700$ nm~ for the G channel, and $\lambda = 700 - 800$ nm~ for the R channel). A map of the H$\alpha$ flux is overlaid in red using a logarithmic scale, revealing the extended gas tail that originates from the high-velocity encounter of ESO137-001 with the intra-cluster medium.}
\label{fig:ram_pressure_anita}
\end{figure}

In the local Universe, the smoking gun of the role of ram pressure
stripping in quenching star
formation is the disturbed gas content of galaxy  cluster members. Radio
surveys at 21 cm revealed the HI deficiency of spiral galaxies in
overdense environments, especially in the vicinity of the cluster core (\citealt{Haynes1985}) and more recently the
molecular gas deficiency has also been reported
(\citealt{Fumagalli2009}, \citealt{Boselli2014}). Furthermore, these
galaxies often show disturbed gas morphologies and tails visible both
in H$\alpha$ (Figure \ref{fig:ram_pressure_anita}) and UV continuum (e.g. \citealt{Gavazzi1985},
\citealt{Fumagalli2014}, \citealt{Fossati2016}).

However, at high redshift ($1 \lesssim z \lesssim 1.5$) there is still no
consensus on the influence of the environment on galaxies' gas content
(\citealt{Aravena2012}, \citealt{Dannerbauer2017},
\citealt{Coogan2018}). This is mainly driven by detections that are either lacking, or limited to the most gas-rich members. 
Galaxies undergoing ram pressure stripping are ideal laboratories to
constrain the efficiency with which gas can be removed and star
formation is quenched. Observations of galaxy clusters and groups at
different redshifts are therefore key to understand how and when the red sequence
of galaxies is assembled in dense environments.

\subsubsection{The need for space-based UV observations}
To study quenching
mechanisms in dense environments and the origin of galaxy red sequence, a space-based instrument with a large field of view ($\sim$ 1 $\times$ 1 arcmin$^2$), exquisite sensitivity, and a spectral coverage $\lambda \sim 90 - 350$ nm is needed. This part of the spectrum includes
the Ly$\alpha$ emission line at redshift $z \lesssim 1.7$. Ly$\alpha$ traces the neutral gas, which is expected to be the
bulk of the material stripped by ram pressure, and is expected to be $>
10\times$ brighter than H$\alpha$ (e.g. \citealt{Scarlata2009}). Mapping the Ly$\alpha$ emission with a wide-field spectrograph would allow us to trace stripped
gas out to large distances from the galaxy ($\sim$ 500 kpc at $z = 1$) and to unveil gas
tidal tails $\gtrsim 10\times$ fainter than the H$\alpha$ ones,
therefore targeting also the less massive cluster members. The detection of other emission lines (e.g. CIII, OI, HeII, MgII) would allow us to constrain the density, temperature, and metallicity of the ionized gas. The comparison of the Ly$\alpha$ and H$\alpha$ fluxes would also give information about the powering mechanisms of Ly$\alpha$ (e.g. shocks, star-formation, cooling), still an unknown issue. 
Are cluster members mainly fast rotators, as expected if their star formation has been quenched by the interaction with the intra-cluster medium? Or are more violent interactions \citep{Toloba2011} responsible for the stripping of their gas? A resolution of $R \sim 4000$, would enable us to probe the kinematics of stars and gas simultaneously, and such questions would be thoroughly addressed. 

The stripped gas is expected to be mainly in the cold atomic phase,
but it could be heated and change phase once it interacts with the hot gas confined within the potential well of the cluster. Complementary multi-frequency observations will allow us 
to reach a comprehensive picture of these phenomena: the molecular phase is detectable
through CO (ALMA), SKA will target the cold HI gas at 21 cm, the hot gas phase is visible in the X-rays
(\textit{Athena}), while the ionized gas phase will be probed by ELT instruments.

\subsection{Supernovae}
\label{subsec:transients}

Core collapse supernovae are the most common form of death for massive stars \citep{li_nearby_2011}. These events are the end-stage of massive stellar evolution, and are responsible for the chemical enrichment of the galaxy. While many supernovae have been discovered and observed, the characteristics of the stars that explode are still a mystery. Observationally it has been established that massive stars ($>8M_\odot$) end their lives as a core-collapse supernova. However, theoretical models have not succeeded in reproducing the observed explosion properties, and more importantly cannot self-consistently explode the star without an artificially induced ignition \citep{burrows_colloquium:_2013}. The progenitors of core-collapse supernovae have been positively identified for a handful of events \citep{smartt_death_2009,smartt_observational_2015}. The progenitor's characteristics are one of the major open questions which must be addressed in order to progress our understanding of the explosion mechanism of stars, as well as the final stages of stellar evolution.
	
The most robust technique for identifying progenitors is with high-resolution images taken before the explosion. This technique has been successfully employed a handful of times using primarily \textit{HST} images taken before the events \citep{smartt_detection_2004,maund_yellow_2011,van_dyk_supernova_2017}. Some core-collapse supernovae seem to originate from red supergaint stars, but there is also an example of a Type IIb originating from a yellow hypergiant \citep{maund_yellow_2011, bersten_type_2012}. Two limitations prevent this technique from significantly advancing the field in the future: first, an image of the region prior to the explosion is needed; second, it is necessary to wait for $5-10$ years until the supernova fades to below the progenitors' luminosity in order to confirm that the star terminally exploded and disappeared. Even when the second criterion is satisfied, some doubt remains as the star could be enshrouded in a high-extinction dust. Given these limitations, pre-explosion imaging is unlikely to increase the number of identified progenitors by orders of magnitude in the future.
	
\begin{figure}[h!]
\centering
\includegraphics[width=0.9\textwidth]{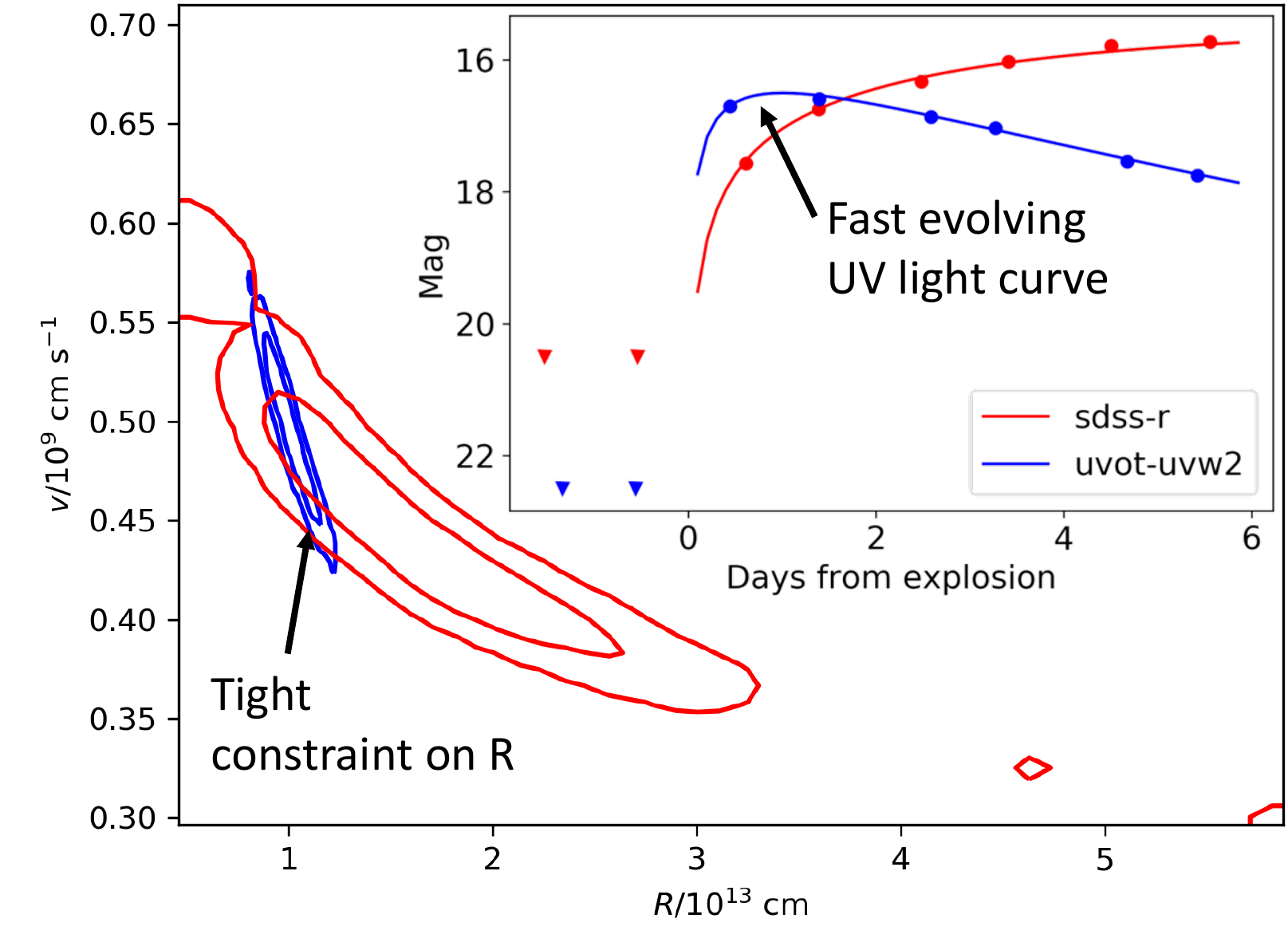}
\caption{Constraints on radius and ejecta velocity (energy per unit mass of the explosion). In the inset we show a simulated light-curve as it would appear in UVOT-UVW2 and SDSS-r. The simulation is of an $R=10^{13}$ cm progenitor observed with $\sim1$ day cadence. Note the faster rise of the UV light-curve. The contours represent 68\% and 95\% of the probability of the fit to the UV and visual bands separately. The UV is clearly more sensitive to the progenitor's radius than visual bands.}
\label{fig:shock_cooling_Adam}
\end{figure}

The solution is to use semi-analytic models which have been developed in recent years \citep{waxman_grb_2007,nakar_early_2010,rabinak_early_2011,sapir_uv/optical_2017}. These models can relate the observed light curve of a core-collapse supernova to the progenitor's radius and the velocity of the ejecta (energy per unit mass of the explosion). These models can be applied to light curves of events which do not have pre-explosion images, and therefore have the potential to increase by orders of magnitude the number of identified progenitors. The hot ejecta ($>10^4$ K) cools rapidly during the first days after the explosion. Because in the UV and optical wavelengths the blackbody spectrum is in the Rayleigh-Jeans regime, the UV light-curve evolves more rapidly than longer optical wavelengths (e.g. R-band). Shock-cooling models can only be applied to data taken during the first few days after explosions, therefore the rapid evolution of the UV light curve is critical for constraining the progenitor's properties. \citet{rubin_exploring_2017} showed that visual-band wavelengths cannot constrain the progenitor's radius meaningfully, but that UV coverage at moderate ($0.5-1$ day cadence) can constrain the progenitor's radius to 20\%. With a significant number of supernovae observed in the UV, it will be possible to measure the progenitor population of core-collapse supernovae. These will provide the initial conditions for explosion models, and will provide benchmarks against which to design and test these simulations.
	
UV spectroscopy will also play a major role in understanding the surface composition of the progenitors. It has been shown that many supernovae have a circumstellar material surrounding the star from before the explosion \citep{gal-yam_wolf-rayet-like_2014,yaron_confined_2017,khazov_flash_2016}. At the moment of first light, this material becomes highly ionized by the tremendous luminosity and temperature of the first photons of the explosions. After a timescale of hours to days the bulk of the supernovae ejecta sweeps up this material. During this window strong, highly ionized emission lines have been observed (e.g. OIII, HeII). These can be related back to the surface composition of the progenitor star in the last period before it exploded. \citet{groh_early-time_2014} showed that the most informative lines are in the UV, e.g. CIV $\lambda1548-1551$, HeII $\lambda1640$, and NVI $\lambda1718$.

\subsubsection{The need for space-based UV observations}
To achieve our science goals we will need photometry of magnitude 20
sources with a cadence of 0.5 -- 1 day. By simultaneously obtaining
the spectra of our targets (with magnitude down to 20) with a
signal-to-noise of 10 we will measure the temporal evolution of the UV flux, and the strength and equivalent widths of the transient emission lines CIV  $\lambda1548-1551$, HeII $\lambda1640$, and NVI $\lambda1718$, to achieve the science goals mentioned above (Figure \ref{fig:shock_cooling_Adam}).

\subsection{Accreting white dwarfs in globular clusters: testing the models of compact binary evolution}
\label{subsec:binaries}

Accreting white dwarfs (WDs), namely binaries in which a WD accretes from a main sequence star or a degenerate companion, are a strategical tool to probe the physical properties of the Universe and to test fundamental physical theories.
The thermonuclear ignition of WDs following the interaction with a binary companion results in Type~Ia supernovae (SNe\,Ia), which are fundamental yardsticks to constrain cosmological distance scales \citep{candles} and the existence of dark energy \citep{riess,perlmutter}. 
Moreover, the most compact systems with orbital periods below one hour are among the brightest known low frequency gravitational wave sources and will be used to verify the performance of the space-based gravitational wave mission \emph{LISA} and to calibrate the detector for future gravitational wave source discoveries \citep{Thomas+2018}.
It is therefore critical to understand the evolution and final fate of accreting WDs.

Currently, there are several significant discrepancies between the predictions of population synthesis models and the observed properties of accreting WDs. In particular (i) the interplay between the angular momentum loss mechanisms driving the evolution of accreting WDs at different orbital period regimes is poorly understood \citep[e.g.][]{Schreiber+2016,Belloni+2020}; (ii) the evolutionary path followed by the most compact systems is unknown \citep[e.g.][]{Green+2018}; (iii) the final fate of accreting WDs and the pathway leading to SN\,Ia explosions is still not clear \citep[see e.g.][for a review]{Maoz+2012}.
Solving these discrepancies is essential before the theoretical models can be sensibly applied to more complex binaries, such as black hole and neutron star binaries, X-ray transients, and SN\,Ia progenitors. 

Accreting WDs are predicted to be numerous in globular clusters ($\simeq 100-200$ per cluster, \citealt{Ivanova+2006,Knigge2012}). Globular clusters (GCs) are therefore a unique laboratory where to carry out statistical binary population synthesis studies for populations of accreting WDs with known distance, metallicity, and age \citep{Knigge2012,Belloni+2016}. By observing populations of accreting WDs in GCs we will explore the following science cases: 

\begin{itemize}
\item 
Constrain the rates of orbital angular momentum losses. 
To test the prediction of the current models, measurements of the
angular momentum loss rates at different orbital periods
($P_\mathrm{orb}$) are needed. However, the long timescale over which
the orbital period changes typically prevents direct measurements from
being obtained. A proxy for the mean mass accretion rate and, in turn, for the angular momentum loss rate is the WD effective temperature ($T_\mathrm{eff}$, \citealt{TowBild, Bildsten+2006}), as it is determined by the compressional heating of the accreted material \citep{Sion1995,Townsley+2004}. While several thousand accreting WDs are known, accurate temperatures are only measured for $\simeq 80$ systems, obtained from ultraviolet \textit{HST} observations \citep{Bildsten+2006,Dean_and_Boris2009, Pala}. In this sample, only systems with a period $70\,\mathrm{min} < P_\mathrm{orb} < 150\,\mathrm{min}$ show an angular momentum loss in agreement with model predictions, while all the others show discrepancies of more than one order of magnitude with the theory (Fig.\,5).

\item The formation of the most compact systems. Two main populations of accreting WDs are known: (1) the systems with main-sequence and brown-dwarf donors (cataclysmic variables, CVs) with orbital periods in the range $60\,$min\,$ \lesssim P_\mathrm{orb} \lesssim 2\,$d and (2) the systems with helium stars or helium-core WDs as donors (AM\,CVn stars) with orbital periods $P_\mathrm{orb} \lesssim 60\,$min.
The formation of AM\,CVn stars is still poorly understood and evolutionary models currently predict different formation scenarios \citep{Nelemans+2010}. Particularly, it has been suggested that AM\,CVn could descend from CVs in which the donor is nuclear evolved, i.e. enriched of CNO processed material. To observationally constrain different models it is key to study the chemical composition (e.g. N/O and N/C ratios) of the donor \citep{Nelemans+2010}. Establishing the fraction of CVs with evolved donor could provide an upper limit on the number of AM CVn that are expected to form through this channel \citep[see e.g.][]{Pala+2020}, thus yielding a valuable observational test for the formation of these compact systems.

\item Accreting WDs and SN\,Ia progenitors. Both CVs and AM\,CVns are intimately connected to the formation of SNe\,Ia. In fact, CVs with nuclear evolved donors undergo a phase of stable hydrogen shell burning during which the WD mass grow (possibly to the Chandrasekhar limit) and AM\,CVns accrete He-rich material from their degenerate donors, potentially triggering a SN\,Ia via the double-detonation mechanism \citep{Bildsten+2007,Fink+2010}. Determining the masses of these systems through a fit of the UV spectrum provides firm statistical constraints for the above scenarios and could finally confirm the potential of these systems as SN\,Ia progenitors. 
Also the rotation rates of accreting WDs are a critical parameter in the pathway leading to SNe\,Ia explosion. Absorption features from the photosphere of the WD accretor, which are only cleanly detected in the UV, can be used to measure the rotational velocity and determine whether the WD is spun up by accretion, allowing the WD to exceed the Chandrasekhar limit without triggering the SN explosion.
\end{itemize}

\begin{figure}[h!]
\centering
\includegraphics[width=0.9\textwidth]{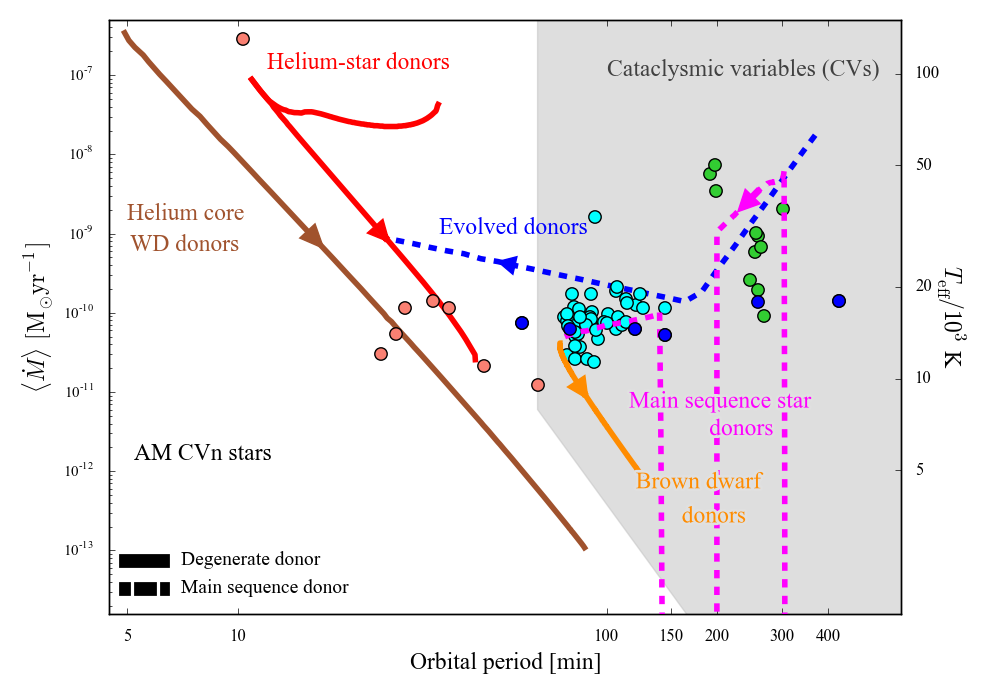}
\caption{Effective temperatures of short-period (cyan) and long-period (green) CV WDs, AM\,CVns (pink) and systems with evolved donors (blue) \citep{Bildsten+2006,Dean_and_Boris2009,Pala}. 
The WD $T_\mathrm{eff}$ (right y-axis) directly translates into a measurement of the average mass accretion rate ($\langle\dot{M}\rangle$, left y-axis), which is a measurement of the angular momentum loss rate in the system. While the observations of short-period CVs (cyan) agree reasonably well with the theoretical evolutionary tracks \citep{Bildsten+2006,Yungelson2008,Nelemans+2010,Pala}, long-period CVs (green), AM\,CVns (pink) and systems with evolved donors (blue) are poorly studied and the few systems observed uncover major discrepancies between observations and models.}
\label{fig:binaries_Anna}
\end{figure}

\subsubsection{The need for space-based UV observations}
The population of accreting WDs in GCs provides a statistically significant sample of systems spanning the whole orbital period distribution and will allow us to carry out stringent tests for the current models of compact binary evolution \citep[e.g.][]{Knigge+2002}. 
Accreting WDs have typical temperatures $T_\mathrm{eff} \gtrsim 10\,000\,\mathrm{K}$ and their spectral energy distribution peaks in the ultraviolet. In other wavelength domains, the emission is dominated by the disc (in the optical) or a boundary layer at the WD surface (in the X-rays). Space-based ultraviolet spectroscopy is therefore the only method to access and fully characterize the physical properties of these stars \citep[e.g][]{Szkody2002,Boris2006,BDPav,Pala}. 
Moreover, the majority of the main sequence and red giants stars are cool (Teff $<$ 7000 K) and their emission peaks in the optical and/or in the near-infrared. For this reason, in the UV wave range GCs appear vastly less crowded than in the optical (see e.g. fig.~1 from \citealt{Knigge+2002}). The proposed UV IFU would provide the possibility to (i) deblend the denser environments and study those sources that would be inaccessible to optical studies owing to crowding and (ii) obtain, in one shot, a spectrum of the detectable UV sources in the field of view.

From the knowledge of the GC distance, the WD $T_\mathrm{eff}$ can be measured via a spectral fit to the ultraviolet data with synthetic atmosphere models thus providing a direct measurement of the angular momentum loss rate in the system \citep{Dean_and_Boris2009}. The spectral fit to ultraviolet data also provides the WD photospheric abundances which reflect the composition of the donor star, via the detection of strong emission and absorption resonance lines of C, N, O and Si (e.g. N{\sc v} 1240\,\AA, C{\sc iv} 1549\,\AA, and O{\sc i} 1150, 1302\,\AA; \citealt{Boris+2003, Morales+2003,Sion+2006}) which are not accessible in the optical. A single ultraviolet spectrum thus provides information on both the WD accretor and the donor star, offering a unique insight into the composition of the donor and the prior evolution of the system.
A spectral fit to ultraviolet data also yields accurate mass determinations, thus providing firm statistical constraints for the double-detonation scenario and the mass growth in CVs, finally confirming the potential of these systems as SN Ia progenitors. 
Finally, rotation rates of accreting WDs have so far been measured only in a handful of systems \citep[e.g.][]{sionetal94-1, longetal04-1}. A resolving power of $\simeq 4000$ is sufficient to measure rotation rates $v \sin i \gtrsim 100\,$km/s, thus providing robust observational constraints on the response of WDs to the accretion of mass and angular momentum.

\section{Comparison with other space instruments}
\label{sec:uniqueness}

A wide-field, IFU-like instrument on board a space telescope optimized for UV observations ($\lambda = 90 - 350$ nm) will allow the community to simultaneously obtain photometric and spectroscopic observations with exquisite sensitivity in a wavelength regime that is not accessible from the ground, making it a unique instrument to tackle the scientific questions presented in Section \ref{sec:sc} as well as many other topics. There are currently only two spectrographs on board \textit{HST} that cover a wavelength range similar to the one that we propose (COS and STIS), one such spectrograph was launched on the stratospheric balloon FIREBall, and two new ones are currently being proposed to be on board of CETUS and the \textit{LUVOIR} telescope (LUMOS). In the following, we discuss why these spectrographs are not suitable to achieve the science cases described in Section \ref{sec:sc}. 

\textbf{\textit{HST}/COS}: the wavelength coverage of this spectrograph ($\lambda = $ 90 - 320 nm) is similar to the one that we propose and it was designed to obtain spectroscopy of faint point-like sources (e.g. stars, quasars) with a resolving power R $\sim$ 1500 -- 24000 \citep{cos2012}. The extremely small field of view of COS ($2.5''$ diameter) however prevents observations of large patches of the sky and the possibility of observing extended objects, a critical requirement for all science cases presented in this document. Some of our science questions overlap with those that motivated COS, but the way we want to address them is substantially different. As an example, the question of whether the cosmic web exists and how baryons from galaxies interact with the surrounding medium can be tackled by COS by studying absorption spectra of intergalactic gas (e.g. Ly$\alpha$ forest, highly ionized absorption lines such as OIV, NV), whereas we propose to address this issue by studing the intergalactic medium in emission, through the detection of Ly$\alpha$ halos extended across hundreds of kpc (and possibly other rest-frame UV emission lines). Similarly, COS is suitable to study white dwarfs, cataclysmic variables, and binary stars in the Milky Way. However the instrument that we propose will open up the new possibility to target these sources in globular clusters, dramatically increasing the statistics and efficiency of the observations.

\textbf{\textit{HST}/STIS}: this spectrograph and imaging camera covers the far- and near-UV wavelength range ($\lambda = $ 115 -- 310 nm, \citealt{stis1998}). Spatially-resolved observations can be performed through slitless spectroscopy, but the resolving power (R $\leq 2500$) is too low to achieve our science goals (R $\gtrsim 4000$ is needed). Furthermore, the low throughput of STIS ($\lesssim$ 10\%) only allows observation of the brightest targets (the limiting magnitude V = 20.6 for an A0V star can be observed with a signal-to-noise ratio of 10 in 1 hour on source) and is not enough to reach the faint surface brightness levels required to achieve our goals (e.g. SB$_{\rm Ly\alpha}\sim10^{-19}-10^{-20}\cgssb$, see Section \ref{subsec:cgm_fab}).

\textbf{FIREBall-2}: this is a multi-object spectrograph operating in the UV and flying on a stratospheric balloon at an altitude of 40 km \citep{Lemaitre2019}. It follows the experiments started with FIREBall-1 aimed at detecting the faint and diffuse emission of the intergalactic medium. It is equipped with a fiber IFU (300, 8" fibers). Part of the scientific motivation for FIREBall overlaps with the one proposed here (Section \ref{subsec:cgm_fab}), but FIREBall only observes a limited wavelength range (200 - 210 nm) and targets the Ly$\alpha$ emission line at $z \sim 0.6 - 0.7$. We propose to cover a  larger wavelength range ($\sim$ 90 -- 350 nm), pushing the observations of Ly$\alpha$ up to z $\sim 1.7$, starting to bridge the local and high-redshift Universe. Furthermore the field of view of FIREBall-2 is not contiguous (i.e. the fibers need to be placed in separate positions) and the spatial ($\sim$ 4") and spectral (R $\sim$ 2000) resolution are not enough to achieve our science goals. The overall throughput of the instrument on FIREBall is $\sim$ 5\% that is not enough to reach the low surface brightness level needed by our science goals. Finally, since this instrument works on a stratospheric balloon, its operations strongly depend on weather conditions and aeronautical rules, and observations can only be performed for a few consecutive nights each time. The possibilities to get the long exposures needed to probe faint emissions are therefore limited. These kind of operational conditions cannot sustain a large demand from the astronomical community, but are key for testing new UV technologies and space-validate sub-components in a zero-gravity environment. The FIREBall team has also proposed to NASA to fund a new mission, ISTOS \citep{Martin2011}, designed at studying the faint UV emission from the circum- and inter-galactic medium leveraging in particular on the development of new detectors with improved capabilities (i.e. higher quantum efficiency and lower noise thanks to the use of EMCCDs).

\textbf{\textit{LUVOIR}/LUMOS} and \textbf{CETUS}: these are both UV
multi-object spectrographs that have been proposed for future
missions. LUMOS will cover the far-UV to visible wavelength range
($\lambda = 100$ -- 1000 nm), is currently under study, and is a
candidate instrument for the future \textit{LUVOIR} space mission
\citep{France2017}. CETUS instead is a Probe Mission Concept that NASA
has selected for consideration \citep{Kendrick2019}. It will enable
parallel observations by the UV multi-object spectrograph ($\lambda
\sim 180 - 350$ nm) and near-UV/far-UV camera ($\lambda 115 - 400$ nm)
which will operate simultaneously but with separate field of
views. Thanks to their large field of view ($2' \times 2'$ LUMOS, and $17' \times 17'$ CETUS) both these instruments will be able to observe hundreds of targets at once. However the field of view covered by LUMOS and CETUS, as opposed to the IFU instrument that we are proposing, will not be contiguous. A 100\% coverage of a large field of view to a high depth is critical to efficiently achieve our science goals, such as contiguously map the extended emission of the cosmic web and galaxy halos, the extended tidal tails around galaxy cluster members, globular clusters that host populations of accreting white dwarfs. A discontiguous field of view drastically reduces the efficiency of the observations and the discovery potential. Additionally, compared to a multi-object spectrograph that can only perform pointed observations, the IFU instrument that we propose will provide thousands of spectra with a single pointing and will therefore have a unique potential for serendipitous discoveries. Finally, LUMOS is designed to observe at the diffraction limit and CETUS will have a spatial resolution $\lesssim 0.3''$, conditions that would prevent us from achieving the extremely low surface brightness levels needed by our science cases.

Additionally the \textbf{MESSIER surveyor}, a small UV ($\lambda = 150 - 1000$ nm) space mission designed to explore the Universe down to very low surface brightness fluxes ($\sim 34 - 37$ mag arcsec$^{-2}$), is under development and has recently received initial funding from the French Space Agency CNES \citep{Valls-Gabaud2017}. Although some of the science drivers of this surveyor overlap with those proposed in Section \ref{sec:sc} (e.g. the detection of the cosmic web in emission), MESSIER will be an imager equipped with a set of filters, whereas we are proposing an IFU instrument that will allow the community to simultaneously obtain photometric and spectroscopic information of the targets. With such an IFU the community could follow-up spectroscopically the most promising targets imaged by facilities like MESSIER. Having access to spectroscopy down to such low surface brightness will enlarge the parameter space that can be explored and greatly increase the discovery potential.

\section{Synergies}
\label{sec:synergies}

In the next two decades all the major facilities are expected to operate at red, IR, and radio wavelengths (e.g. \textit{JWST}, ELT, SKA). A window on the UV will therefore offer a complementary view at shorter wavelengths and will have strong synergies with these facilities. It will be critical to conduct follow-up observations in the UV regime, at a time when \textit{HST} will not be operational anymore. An IFU-like instrument operating in the UV will also provide targets that can then be followed-up with the ELT (e.g. the gas stripped from galaxy cluster members observed in Ly$\alpha$ can be observed in [OII] and H$\alpha$ with ELT/HARMONI providing information on the ionized gas). Furthermore most of the science questions presented here have synergies with future major facilities such as SKA at longer wavelengths (e.g. to detect the HI gas stripped in cluster members and the cosmic web around galaxies) and will benefit from targets identified by \textit{Euclid} and the Rubin Observatory in large field imaging. There are also synergies with ALMA, should it still be operational (e.g. the gas stripped by cluster members might be detected both in the neutral phase through Ly$\alpha$ emission with our proposed instrument and in the molecular phase with ALMA). Finally synergies with \textit{Athena} and \textit{LISA} will enhance the discovery potential (e.g. \textit{Athena} will unveil the hot gas phase stripped by cluster members and present in the circumgalactic medium visible in the X-rays, while \textit{LISA} will detect the gravitational waves produced by the most compact accreting white dwarfs with short orbital periods).

\section{Technical concept}
\label{sec:performances}
In this Section, we present some key design areas for a system compliant with the proposed science theme. We notice that a dedicated space mission is not strictly required and an instrument hosted on another future spacecraft may suffice. Finally, we highlight that a system study has not been performed, as this goes beyond the scope of the proposal. 
Critical technologies are highlighted via a brainstorming process and discussions with relevant experts.
Given the main characteristics of the telescope (Table \ref{tab:tech}), the dedicated space mission fits an ESA M-size type. 

\subsection{Type of Telescope and Instrument}

Current space-based facilities are optimized to detect point-like sources such as stars or distant galaxies, rather than extended low surface brightness targets \citep{Trujillo2016}. To reach the surface brightness of $\sim 10^{-19}$ erg s$^{-1}$ cm$^{-2}$ arcsec$^{-2}$ required by our science cases, a telescope with a diameter of $\sim$ 0.9 -- 1.1 m and a relatively fast optics (F-number $\sim$ F/5 -- F/15) is needed \citep{Valls-Gabaud2017}. Ideally, the F-number should have even faster values. That would require technologies with currently a very low technology readiness level (TRL), such as curved detectors.

A UV integral field unit (IFU) is the only available option to achieve the science goals of the previous Sections. Part of the technology that has already been developed for the ground-based Multi Unit Spectroscopic Explorer (MUSE) on the Very Large Telescope (VLT) and that is currently being upgraded for BlueMUSE ($\lambda = $ 350 -- 600 nm, \citealt{Richard2019}) can be adapted for this project. Insights on the performance of IFUs in space can also be obtained from \textit{JWST}/NIRSpec \citep{Deshpande2018}, when launched, although NIRSpec will operate at IR wavelengths. 

The proposed IFU extends the capabilities of such type of instruments toward bluer wavelengths, $\lambda = $ 90 -- 350~nm, not observable from the ground, targeting different science cases. It has a single instrumental setup (i.e. fixed spectral and spatial setup), which simplifies the overall fore-optics. A rapid response mode is necessary to carry out observations of transients (Section \ref{subsec:transients}). We summarize in Table \ref{tab:tech} the main characteristics of the instrument, with their reasonable range.

\begin{table}
\centering
\caption{Main (range of) parameters of the telescope and  instrument.}
\renewcommand{\arraystretch}{1.3}
\label{tab:tech}
\begin{tabular}{c|c}
\toprule
Aperture & 0.9 - 1.1 m \\
\hline
F-number & F/5 - F/15 \\ 
\hline
Field of view & 1 $\times$ 1 arcmin$^2$ \\
\hline 
\hline
Wavelength range & 90 -- 350 nm (preferred) \\ 
                 & 100 -- 300 nm (minimum requirement) \\
\hline
Spectral resolution & average R = 4000 \\
\hline 
Spectral sampling & 0.5 \AA~ per spectral bin \\
\hline 
Spatial sampling & 0.5" $\times$ 1" per spaxel \\
\hline
Spatial resolution (FWHM) &  2" -- 3" \\
\hline 
Throughtput & $\gtrsim$ 25\% over the whole wavelength range \\
\bottomrule
\end{tabular}
\end{table}

\subsection{Mass}
Due to the complexity and extent of the optics, IFUs tend to be heavy instruments, an important constraint for a space mission. The mass will be one of the main design drivers. Section \ref{subsec:detectors} proposes a possible way to greatly simplify the optics. That is a solution with low TRL and requiring further development.

\subsection{Thermal Control}
Another design driver is given by the thermal environment, as it is often the case for optics in space. On one hand, for the detectors, cryogenic temperatures are required to maximize the performance. On the other hand, it is preferable to perform UV observations with a temperature of the optics above 260~K to prevent absorption on the reflective surfaces, given that UV systems are sensitive even to mono-layers of contaminants \citep{LuvoirRep}.  An operating temperature in the range 270~K -- 290~K is recommended because it reduces the risks linked to launch shocks and facilitates mission development as testing, alignment, and mirrors polishing and figuring can be performed on the ground without cooling facilities, with a smaller number of iterations and eventually lower complexity.

Wavefront stability requires the temperature of the optics to be controlled within $10^{-2}$ K. A sun-synchronous orbit would be preferable to avoid eclipses and operate in a more stable environment. However, this is not a strict constraint and UV observations have extensively been performed by other telescopes (e.g. \textit{HST}) on different orbits.

\subsection{Vibrations Control}
The wavefront stability also calls for isolation and damping of
vibrations sources. Dedicated precision mechanisms on the optics or
micronewton thrusters, following the heritage of Gara and LISA Pathfinder \citep{LTP}, are potential disturbance avoidance solutions \citep{NASAmicrovib}. In fact, they remove the need of reaction wheels, which are strong sources of disturbance also during operation.

\subsection{Materials}
In line with decades of warm optics design, the traditional choices
for the material of the mirrors are Zerodur\texttrademark~ and
ULE\texttrademark~ thanks to their high stability. Silicon Carbide
(SiC) is also an option. In fact, while it poses some challenges on
the surface finishing and requires more stringent constraints on the
thermal control, it has significant advantages in terms of
controllability thanks to its superior strength. A set of actuators
(e.g. piezo-stacks) annealed in the SiC substrate, allows wavefront errors to be compensated in several load conditions, including on-ground testing in 1-g. 

Space-based instruments that are currently operational at UV wavelengths have a throughtput $< 25$\% in the wavelength range $\lambda =$ 200 -- 400 nm~ (e.g. WFC3 UVIS filters on board \textit{HST}). To achieve our goals a throughtput $\gtrsim 25$\% is needed. Such performance is currently hindered by the decrement of reflectivity below 110 nm of the typical mirror coatings (e.g. silver and gold), \citep{LuvoirRep}. Aluminium has in principle a good reflectivity down to wavelengths of $\sim$ 100 nm, but it has the known tendency to oxidation. There are however promising studies on this subject (\citealt{Balasubramanian2017}, \citealt{Quijada}) that point to the use of protective layers to prevent aluminum oxidation from degrading the performance. 

\subsection{Detectors}
\label{subsec:detectors}
To achieve the proposed science goals we need to reach very low
surface brightness objects. As an example, to detect the Ly$\alpha$
emission from the circumgalactic medium we need to observe SB$_{\rm
  Ly\alpha}\sim10^{-19}\cgssb$. Given the characteristics of the
proposed telescope and instrument (Table \ref{tab:tech}), this
translates to a detection threshold of $\sim 0.2 - 1$ e$^-$
pixel$^{-1}$ hr$^{-1}$. Currently the main limitation to reach such
low detection thresholds is given by the readout noise of the
detectors ($\sim$ 2 -- 3 e$^-$ for both CMOS and CCDs). However, there
are ongoing studies focusing on enhancing UV detector performances and
reducing the readout noise. On the CCD side, EMCCDs that use avalanche
gain on the readout are currently available. This technology has been tested in sub-orbital flights on FIREBall (\citealt{Hamden2016,Hamden2019}). 

In the CMOS detector market, 8k $\times$ 8k CMOS devices exist already, although they are not optimized yet for the UV and they still have a relatively high readout noise. A study to develop a CMOS detector with readout noise below 0.15 e$^-$ is currently in progress with results from the simulation of the silicon circuit showing single-photon sensitivity (\citealt{Stefanov2020}).  Another promising technological development is in black silicon nano-structures with self induced junctions, allowing effective QE above 100\% without external amplification in the UV wavelength range (\citealt{Garin2019}). A combination of these technologies would lead to a powerful new detector optimized for low-noise UV observations.

An alternative to building an optical IFU with traditional detectors is the use of an imaging camera together with a "spectrometer on a chip" detector based on MKID technology. These detectors measure the energy of every photon that hits the detector, and are already in use for the DARKNESS spectrometer \citep{Meeker2018}. MKIDs have no dark current or readout noise in the traditional sense, as they measure thousands to millions of quasiparticles for every photon that hits. However, they are subject to quasiparticle fluctuations, limiting the precision with which the energy of the photons can be measured and in turn the achievable spectral resolution. In fact, the spectral resolution that they can currently achieve is only R < 100, but theoretically it could be much higher and there are already proposals for using this technology in space \citep{Rausher2016}. This technology however trades decreased optics complexity for increased cryogenic complexity, as MKIDs require a superconducting layer and therefore extreme cryogenic temperatures ($\approx$1~K).

\section{Conclusions}

An entire new window on the Universe is opened by exploring its UV-range with high spatial and spectral resolution. We have showcased several different science questions to be addressed in the time frame of Voyage 2050, and highlighted how our approach will ensure ample opportunities for serendipitous discoveries. 

The ideal instrument to answer the proposed scientific questions is a UV-optimized, wide-field IFU. It could be hosted by a future telescope or alternatively by a new M-size mission and it will be the only UV IFU in space. Designing and building such a complex system will encourage the technological development of several areas, including material science, data acquisition and reduction, UV detector technology, and micro-vibrations mitigation.

Finally, by focusing on the UV wavelength range after and during an epoch when the largest available facilities have been operating at red and IR wavelengths (e.g. \textit{JWST}, ELT), this instrument will open up a new discovery space and will be suitable to work in synergy with the main available observatories now in development such as \textit{Athena}, the ELT, and SKA.

\begin{acknowledgements}
We are grateful to J. Kosmalski, J. Spyromilio, Chian-Chou Chen,
and F. Lelli for useful discussions and inputs.
\end{acknowledgements}

\section*{Conflict of interest}
The authors declare that they have no conflict of interest.

\bibliographystyle{spbasic}      
\bibliography{draft_white_paper}   

%
%

\end{document}